\documentclass{article}
\pdfoutput=1

\PassOptionsToPackage{table,dvipsnames,xcdraw}{xcolor}
\PassOptionsToPackage{format=default,font=small,labelfont=it}{caption}

\usepackage[preprint]{neurips_2026}

\usepackage[utf8]{inputenc}
\usepackage[T1]{fontenc}
\usepackage{microtype}
\usepackage{amsmath}
\usepackage{amsfonts}
\usepackage{amssymb}
\usepackage{booktabs}
\usepackage{multirow}
\usepackage{tabularx}
\usepackage{graphicx}
\usepackage{listings}
\usepackage{xcolor}
\usepackage{caption}
\usepackage{url}
\usepackage{array}
\usepackage{enumitem}

\usepackage{tikz}
\usetikzlibrary{positioning,calc,arrows.meta}

\usepackage[colorlinks=true,
            linkcolor=blue!60!black,
            citecolor=blue!60!black,
            urlcolor=blue!60!black]{hyperref}
\usepackage{cleveref}

\usepackage{fancyhdr}
\renewcommand{\headrulewidth}{0.4pt}

\graphicspath{{assets/}{figs/}}

\definecolor{codegreen}{rgb}{0,0.6,0}
\definecolor{codepurple}{rgb}{0.58,0,0.82}
\definecolor{backcolour}{rgb}{0.97,0.97,0.97}

\lstdefinestyle{pyshort}{
    backgroundcolor=\color{backcolour},
    commentstyle=\color{codegreen},
    keywordstyle=\color{blue!70!black},
    stringstyle=\color{codepurple},
    basicstyle=\ttfamily\scriptsize,
    breaklines=true,
    keepspaces=true,
    showstringspaces=false,
    language=Python,
    frame=single,
    framesep=3pt,
    xleftmargin=4pt,
    rulecolor=\color{black!20},
    aboveskip=3pt,
    belowskip=3pt,
}
\newcommand{\fv}{\texttt{fast-vollib}}
\newcommand{\fvrepo}{\href{https://github.com/raeidsaqur/fast-vollib}{\texttt{fast-vollib}}}
\newcommand{\pv}{\texttt{py\_vollib}}
\newcommand{\pvv}{\texttt{py\_vollib\_vectorized}}

\title{\raisebox{-0.38\height}{\includegraphics[width=0.08\textwidth]{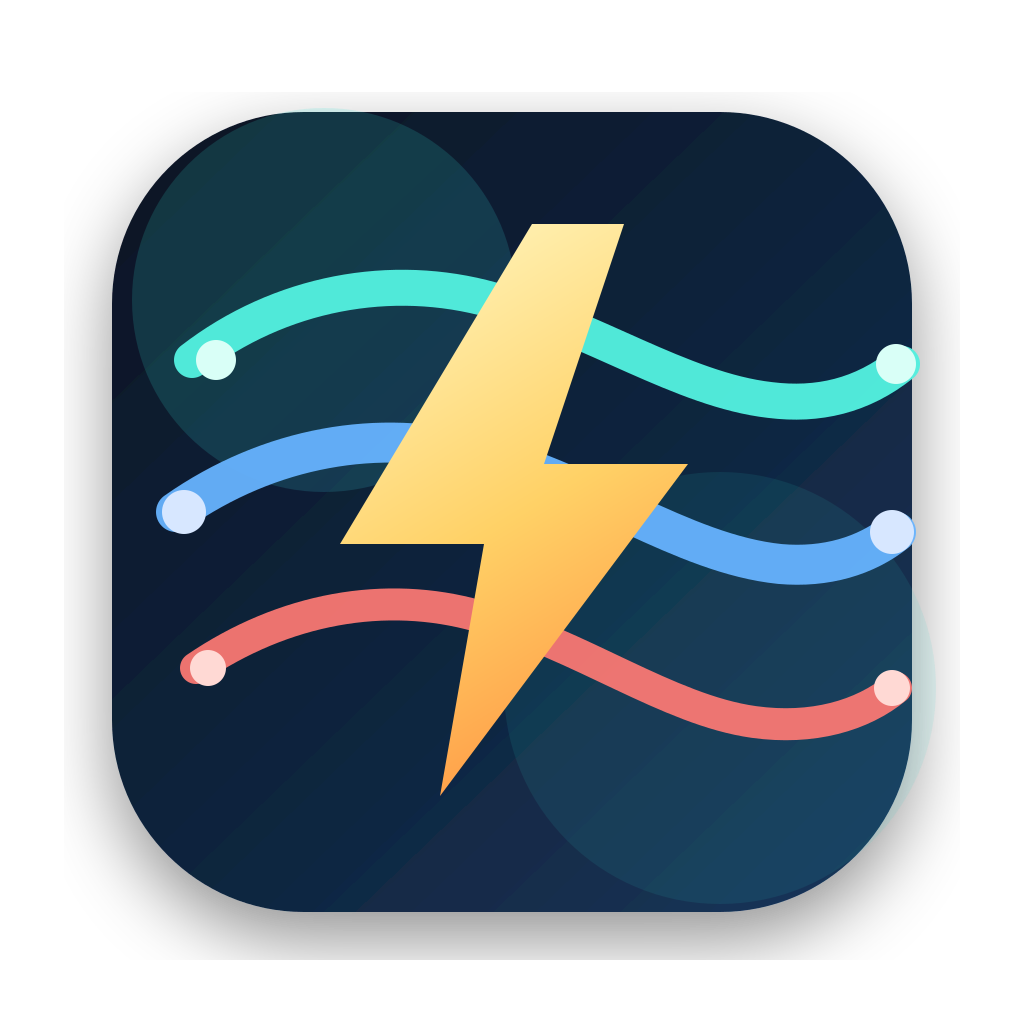}}\,%
\textbf{\fv} \\ A Fast Implied Volatility Library for Python with PyTorch, JAX, and CUDA Fused-Kernel Backends}

\author{%
  Raeid Saqur\\
  Mathematical Institute, University of Oxford\\
  Vector Institute\\
  \texttt{raeid.saqur@maths.ox.ac.uk}
}

\begin{document}
\maketitle

\begin{abstract}
\noindent
We present \fv, an open-source Python library that provides high-performance
European option pricing, implied volatility (IV) computation, and Greeks under
the Black-76, Black-Scholes, and Black-Scholes-Merton models.
The library is designed as a drop-in alternative to the de-facto-standard \pv\ and \pvv\ packages, with pluggable PyTorch and JAX execution backends, a CUDA fused-kernel Triton contribution for batched IV workloads, and a compatibility-first public API.
In addition to a vectorized Halley-method IV solver, \fv\ ships an experimental, fully-vectorized implementation of J\"{a}ckel's ``Let's Be Rational'' (LBR) algorithm with NumPy/Numba, \texttt{torch.compile}, JAX, and Triton single-pass GPU kernels for batched option chains.
This note announces the library, describes its public API surface, and points to its source, documentation, and packaging artifacts. Library public at: \fvrepo.
\end{abstract}

\section{Introduction}\label{sec:intro}
Succinctly, implied volatility (IV) is the inverse of the Black-Scholes pricing map: it is the unique $\sigma$ that equates a model price to an observed market price, and it
is the universal language in which financial options are quoted, hedged, and
risk-managed~\citep{black1973pricing,merton1973theory,black1976pricing}.
Computing IV is a per-quote root-finding problem that, in modern workloads,
must be performed on millions of contracts -- entire option chains across
strikes, maturities, and time -- and increasingly inside differentiable
pipelines for calibration, deep hedging, and neural surface
construction~\citep{buehler2019deep,horvath2021deep}.

The two standard numerical approaches are (i) Newton- or Halley-style iteration
from a Brenner-Subrahmanyam-type initial guess~\citep{manaster1982iterative,
brenner1988new,corrado1996efficient}, which is simple but loses digits near the
wings; and (ii) Peter J\"{a}ckel's seminal ``Let's Be Rational'' (LBR)
algorithm~\citep{jackel2015let}, which combines a four-region rational initial
guess with a Householder(3) iteration in a normalised Black coordinate system
and reaches machine precision in essentially two iterations. The reference LBR
implementation is a scalar C library; the widely used Python wrappers --
\pv~\citep{pyvollib} and \pvv~\citep{pyvollibvectorized} -- delegate to that
extension, remain CPU-only and/or provide acceleration via \texttt{numba}.

\fv\ provides a modern, multi-backend implementation of these primitives in
pure Python with optional PyTorch and JAX acceleration, plus an experimental
GPU-fused J\"{a}ckel solver. The library is meant as a (i) practical
batched IV computation tool for ML/AI quant pipelines, and (ii) a
\pv-compatible drop-in for existing Python codebases.

The present note serves as a concise, citable reference for the library and
summarises the public API, implementation structure, and available resources.

\section{Library Design}\label{sec:libdesign}

fast-vollib is structured as a thin layered stack
(Figure~\ref{fig:architecture}). The public API exposes
\texttt{py\_vollib}-compatible functions; calls are dispatched at runtime
to one of three computation backends (NumPy, PyTorch, JAX) with optional
Numba and Triton acceleration on the relevant backend. Two implied
volatility solvers --- a reference-grade J\"{a}ckel path and a
throughput-oriented Halley path --- share preprocessing, postprocessing,
and the public API and are selected by a single \texttt{solver} keyword.

\begin{figure}[t]
\centering
\resizebox{0.92\textwidth}{!}{%
\begin{tikzpicture}[
    box/.style={draw, rounded corners, minimum width=2.8cm, minimum height=0.7cm, align=center, font=\small},
    thickbox/.style={box, line width=1.0pt},
    arr/.style={->, thick}
]
\node[thickbox, fill=blue!10, minimum width=9.0cm] (api) at (0, 3.5)
    {\texttt{fast\_vollib} public API \quad
     (\texttt{fast\_black\_scholes}, \texttt{fast\_implied\_volatility},
     \texttt{get\_all\_greeks}, \ldots)};

\node[box, fill=orange!8, minimum width=3.0cm] (compat) at (-5.8, 2.4)
    {\texttt{compat}\\monkey-patch helpers};

\node[thickbox, fill=yellow!15, minimum width=4.2cm] (jackel) at (-2.0, 2.4)
    {J\"{a}ckel path\\(reference-grade)};
\node[thickbox, fill=yellow!15, minimum width=4.2cm] (halley) at (2.5, 2.4)
    {Halley$\times 8$ path\\(throughput)};

\node[thickbox, fill=gray!15, minimum width=9.0cm] (disp) at (0, 1.3)
    {Backend dispatcher \quad
     \texttt{backend="auto"|"numpy"|"torch"|"jax"}};

\node[thickbox, fill=orange!15] (np) at (-3.2, 0.1)
    {NumPy\\backend};
\node[thickbox, fill=green!15] (pt) at (0, 0.1)
    {PyTorch\\(CPU/CUDA)};
\node[thickbox, fill=purple!15] (jax) at (3.2, 0.1)
    {JAX\\backend};

\node[box, fill=green!8, minimum width=4.5cm] (tc) at (0, -1.1)
    {Optional acceleration:\\Numba (CPU), Triton (GPU)};

\draw[arr] (api) -- (jackel);
\draw[arr] (api) -- (halley);
\draw[arr] (jackel) -- (disp);
\draw[arr] (halley) -- (disp);
\draw[arr] (disp) -- (np);
\draw[arr] (disp) -- (pt);
\draw[arr] (disp) -- (jax);
\draw[arr] (pt) -- (tc);
\draw[arr] (np.south) -- ++(0,-0.6) -| (tc.west);
\draw[arr] (compat) -- (api);
\end{tikzpicture}
}
\caption{fast-vollib architecture. The public API routes through either
the J\"{a}ckel path (reference-grade) or the Halley$\times 8$ path
(throughput) to a NumPy, PyTorch, or JAX backend. Optional
Numba/Triton acceleration applies on the relevant backend. The
compatibility layer monkey-patches existing
\texttt{py\_vollib}/\texttt{py\_vollib\_vectorized} call sites without
caller changes.}
\label{fig:architecture}
\end{figure}

\subsection{Public API}\label{ssec:api}

fast-vollib preserves the \texttt{py\_vollib\_vectorized}
\cite{pyvollib} calling convention.
Pricing is exposed through \texttt{fast\_black\_scholes\_\allowbreak merton};
inversion through \texttt{fast\_implied\_\allowbreak volatility} (BSM) and
\texttt{fast\_implied\_\allowbreak volatility\_\allowbreak black} (Black-76);
Greeks through \texttt{get\_all\_\allowbreak greeks}.
All entry points accept broadcastable arrays and a \texttt{backend}
keyword, and return NumPy arrays, native tensors, or DataFrames.
The compatibility helper
\texttt{patch\_py\_vollib\_\allowbreak vectorized()} replaces the upstream
namespace at import time, so existing call sites can switch library
implementation without changing caller code.

\subsection{Dual-Solver Design}\label{ssec:dual_solver}

fast-vollib provides two IV solvers that share preprocessing,
postprocessing, broadcasting, and the public API but differ in the
core iteration:

\begin{itemize}[leftmargin=*]
    \item \textbf{J\"{a}ckel path} (\texttt{solver="jackel"}). A
    direct vectorized implementation of LBR~\cite{jackel2015let}: a
    4-region normalised-Black evaluation
    (\texttt{erfcx} with asymptotic, small-$t$, and CDF-direct
    branches), a 4-branch rational initial guess
    (Hermite cubic on the inflection-bounded interior, rational
    cubic on the wings), a 3-branch transformed objective
    (log-space lower, direct middle, log-complementary upper), and
    a Householder(3) iteration. Two iterations suffice for machine
    precision; we run three on the GPU path as a safety margin.
    \item \textbf{Halley$\times 8$ path} (\texttt{solver="halley"}).
    A Brenner--Subrahmanyam~\cite{brenner1988new} ATM initial guess
    followed by 8 Halley iterations on the direct BSM residual,
    with a bisection fallback on the (typically $< 0.5\%$) rows
    that fail a convergence check. The fixed iteration count makes
    the loop body statically analysable by
    \texttt{torch.compile}~\cite{pytorch2} and TorchInductor; the
    full iteration is fused into a small set of GPU kernels.
\end{itemize}

The contribution of the dual-solver design is not novelty of either
iteration --- both are textbook --- but the explicit, measurable
trade-off under one API. A practitioner who needs reference-grade
labels picks \texttt{solver="jackel"}; a label-pipeline user who
trains in float32 and only needs $\sim\!8$ digits picks
\texttt{solver="halley"}; the choice is in caller code, not a fork
of the library.

\subsection{Implementation Substrate}\label{ssec:substrate}

The library is fully vectorized. All inner kernels operate on entire
arrays; \texttt{where}-style branch selection avoids data-dependent
control flow; the normal CDF is evaluated through the complementary
error function (\texttt{scipy.special.ndtr},
\texttt{torch.special.erfc}, \texttt{jax.scipy.special.erfc}) to
preserve precision for deep-OTM contracts. When Numba is available,
the J\"{a}ckel path's hot loops are JIT-compiled with parallel
execution over batches; when Triton~\cite{tillet2019triton} is
available, both solvers fuse their iteration into a single GPU kernel
with all intermediates kept in registers. These kernel-design choices
are invisible to callers but central to the throughput envelope.

The Triton kernels are a material part of the contribution.  The
J\"{a}ckel kernel fuses preprocessing, the six normalised-Black
evaluations needed for the boundary points $(s_l,s_c,s_h)$, the
Hermite initial guess, and three Householder(3) corrections into one
launch; each option is read from HBM once and the transformed-objective
branches are selected in registers.  The Halley kernel similarly
fuses loop-invariant hoisting, eight fixed Halley sweeps, and the
convergence check; only the small nonconverged mask is sent to the
bisection fallback.  This fusion is why the API can expose both an
audit-grade path and a throughput path without changing the caller's
data layout.

\subsection{Black-76 Convention}\label{ssec:black76_api}

Because correct Black-76 inversion is a common source of convention
errors, fast-vollib exposes Black-76 explicitly rather than only
through a BSM reduction. The function
\texttt{fast\_implied\_volatility\_black(mid, F, K, r, T, flag)}
treats the input \texttt{mid} as a \emph{discounted} Black-76 price
and uses the same \texttt{r} for the discount factor and the
inversion. This avoids the convention mismatches that arise when a
BSM-with-zero-rate inverter is applied to discounted Black-76 mids.

\section{Capabilities}\label{sec:cap}

\fv\ is structured around a small public API that mirrors \pvv\ naming
conventions while exposing modern execution backends. The high-level
capabilities are:

\begin{itemize}[leftmargin=*, itemsep=1pt, parsep=0pt, topsep=2pt]
\item \textbf{Pricing models.} Black-76 (\texttt{fast\_black}),
      Black-Scholes (\texttt{fast\_black\_scholes}), and
      Black-Scholes-Merton with continuous dividend yield
      (\texttt{fast\_black\_scholes\_merton}).
\item \textbf{Implied volatility.}~\texttt{fast\_implied\_volatility} for BSM
      quotes and \texttt{fast\_implied\_volatility\_black} for futures-style
      Black-76 quotes. The default solver uses a vectorized Halley iteration
      with bisection fallback. An experimental \texttt{fast\_vollib.jackel}
      module provides LBR-style IV via NumPy+Numba
      (\texttt{jackel\_iv\_black}), \texttt{torch.compile}
      (\texttt{jackel\_iv\_black\_torch}), JAX (\texttt{jackel\_iv\_black\_jax}),
      and a single-pass Triton kernel
      (\texttt{jackel\_iv\_triton}).
\item \textbf{Greeks.} \texttt{vectorized\_delta},
      \texttt{vectorized\_gamma}, \texttt{vectorized\_theta},
      \texttt{vectorized\_rho}, \texttt{vectorized\_vega}, plus
      \texttt{get\_all\_greeks} which evaluates all five Greeks in a single
      backend call to avoid redundant $d_1$/$d_2$ work.
\item \textbf{Vectorization and batching.} All pricing, IV, and Greek
      functions accept array-like inputs (NumPy arrays, lists, scalars,
      pandas Series) and broadcast the arguments using NumPy broadcasting
      rules. A DataFrame helper, \texttt{price\_dataframe}, prices, inverts,
      and computes Greeks for every row of a \texttt{pandas.DataFrame} given
      column names.
\item \textbf{Option flag conventions.} Inputs follow the \pv\ convention:
      \texttt{"c"} for call and \texttt{"p"} for put, accepted as either a
      scalar string or an array of strings.
\item \textbf{Pluggable backends.} A single \texttt{backend} keyword
      (\texttt{"auto"}, \texttt{"numpy"}, \texttt{"torch"}, \texttt{"jax"})
      selects the execution engine. Auto-resolution prefers CUDA-capable
      PyTorch over JAX over NumPy and can be overridden via
      \texttt{set\_backend} or the \texttt{FAST\_VOLLIB\_BACKEND} environment
      variable.
\item \textbf{Output containers.} A \texttt{return\_as} keyword returns
      \texttt{pandas.DataFrame} (default), \texttt{pandas.Series},
      \texttt{numpy.ndarray}, \texttt{dict}, or JSON; \texttt{return\_native}
      preserves backend-native tensors.
\item \textbf{Drop-in compatibility.} \texttt{patch\_py\_vollib()} and
      \texttt{patch\_py\_vollib\_vectorized()} monkey-patch the upstream
      namespaces so existing user code transparently dispatches to \fv.
\item \textbf{Differentiable IV.} An optional autograd-friendly entry point,
      \texttt{implied\_volatility\_autograd}, is exposed when PyTorch is
      installed.
\end{itemize}

\subsection{Installation and Quick Use}

\begin{lstlisting}
pip install fast-vollib                   # core (NumPy backend)
pip install "fast-vollib[torch]"          # + PyTorch backend
pip install "fast-vollib[jax]"            # + JAX backend
pip install "fast-vollib[torch,jax]"      # both backends
\end{lstlisting}

\noindent A minimal IV inversion on a small batch:

\begin{lstlisting}
import numpy as np
import fast_vollib

prices = fast_vollib.fast_black_scholes(
    flag=np.array(["c", "c", "p"]),
    S=100.0, K=np.array([95, 100, 105]),
    t=0.25, r=0.05, sigma=0.20,
    return_as="numpy",
)

iv = fast_vollib.fast_implied_volatility(
    price=prices,
    S=100.0, K=np.array([95, 100, 105]),
    t=0.25, r=0.05,
    flag=np.array(["c", "c", "p"]),
    return_as="numpy",
)
\end{lstlisting}

\subsection{API surface}

Table~\ref{tab:api} summarises the major public entry points; full signatures
and parameter semantics are documented at the project documentation site.

\begin{table}[!htb]
\centering\small
\renewcommand{\arraystretch}{1.05}
\begin{tabularx}{\linewidth}{@{}lX@{}}
\toprule
\textbf{Symbol} & \textbf{Purpose} \\
\midrule
\texttt{fast\_black} & Black-76 pricing on forward $F$ \\
\texttt{fast\_black\_scholes} & Black-Scholes pricing on spot $S$ \\
\texttt{fast\_black\_scholes\_merton} & BSM pricing with dividend yield $q$ \\
\texttt{fast\_implied\_volatility} & Vectorized IV (BSM) -- Halley + bisection \\
\texttt{fast\_implied\_volatility\_black} & Vectorized IV (Black-76) \\
\texttt{vectorized\_\{delta,gamma,theta,rho,vega\}} & Individual Greeks \\
\texttt{get\_all\_greeks} & All five Greeks in one batched call \\
\texttt{price\_dataframe} & End-to-end pricing/IV/Greeks for a DataFrame \\
\texttt{patch\_py\_vollib} & Drop-in patch of \pv\ namespace \\
\texttt{patch\_py\_vollib\_vectorized} & Drop-in patch of \pvv\ namespace \\
\texttt{set\_backend} / \texttt{get\_backend} & Backend selection \\
\texttt{jackel.jackel\_iv\_black} & LBR IV (NumPy + Numba) \\
\texttt{jackel.jackel\_iv\_black\_torch} & LBR IV (\texttt{torch.compile}) \\
\texttt{jackel.jackel\_iv\_black\_jax} & LBR IV (\texttt{jax.jit}) \\
\texttt{jackel.jackel\_iv\_triton} & LBR IV (single-pass Triton kernel) \\
\bottomrule
\end{tabularx}
\vspace{0.5em}
\caption{Major public entry points of \fv. All non-J\"{a}ckel functions accept
\texttt{flag}, \texttt{return\_as}, \texttt{dtype}, \texttt{backend}, and
\texttt{return\_native} keywords.}
\label{tab:api}
\end{table}

\section{Design and Implementation}\label{sec:design}

The library is organised around a single dispatch layer (\texttt{config.py})
and three backend modules (\texttt{backends/\{numpy,torch,jax\}.py}) that all
implement a common interface (\texttt{price\_*}, \texttt{greeks},
\texttt{implied\_volatility}). Public-facing functions share a uniform
preprocessing pipeline: flags are normalised, arrays are broadcast and
validated for NaN/inf, the requested backend is resolved and invoked, and the
result is formatted into the requested container. A backend-parity test suite
checks consistency across the NumPy, PyTorch, and JAX implementations.

\begin{figure}[t]
\centering
\includegraphics[width=0.85\linewidth]{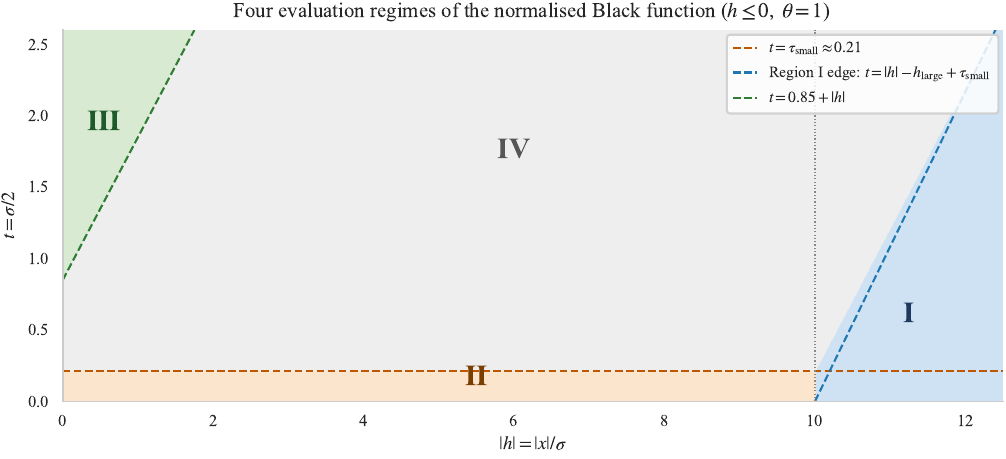}
\caption{The four rational-initial-guess regimes of the normalised Black
function used by the LBR algorithm. Each regime uses a separate rational
approximation before Householder refinement.}
\label{fig:jackel-regimes}
\end{figure}

The Halley-style vectorized IV solver is implemented elementwise in pure
array ops so that \texttt{torch.compile} and \texttt{jax.jit} can fuse the
update body across iterations. The J\"{a}ckel module reimplements LBR's
normalised Black evaluation, four-branch rational initial guess (Fig.~\ref{fig:jackel-regimes}), three-branch transformed objective, and Householder(3) iteration as elementwise array ops,
and additionally provides a single-pass Triton kernel that keeps all
intermediate state in registers~\citep{tillet2019triton}.

\paragraph{Upstream CPU baseline.} Table~\ref{tab:pvv-baseline} reproduces
the \pvv\ benchmarking page's IV-inversion numbers as a historical Python
CPU reference~\citep{pyvollibvectorized}.

\begin{table}[!htb]
\centering\small
\setlength{\tabcolsep}{4pt}
\begin{tabular}{rccccc}
\toprule
\textbf{Contracts} & \texttt{pandas.apply} & \texttt{for}-loop & \texttt{iterrows} & list-comp & \pvv \\
\midrule
10      & 0.037\,s & 0.023\,s & 0.008\,s & 0.023\,s & 0.004\,s \\
100     & 0.069\,s & 0.226\,s & 0.078\,s & 0.225\,s & 0.002\,s \\
1{,}000   & 0.652\,s & 2.322\,s & 0.797\,s & 2.291\,s & 0.003\,s \\
10{,}000  & 6.618\,s & 23.350\,s & 8.186\,s & 23.146\,s & 0.011\,s \\
100{,}000 & 60\,s cap & 60\,s cap & 60\,s cap & 60\,s cap & 0.095\,s \\
\bottomrule
\end{tabular}
\vspace{0.5em}
\caption{Upstream \pvv\ benchmark numbers (wall-clock IV inversion), as
published by the \pvv\ documentation. Reproduced for reference; not measured by
this paper.}
\label{tab:pvv-baseline}
\end{table}

\begin{figure}[t]
\centering
\includegraphics[width=0.95\linewidth]{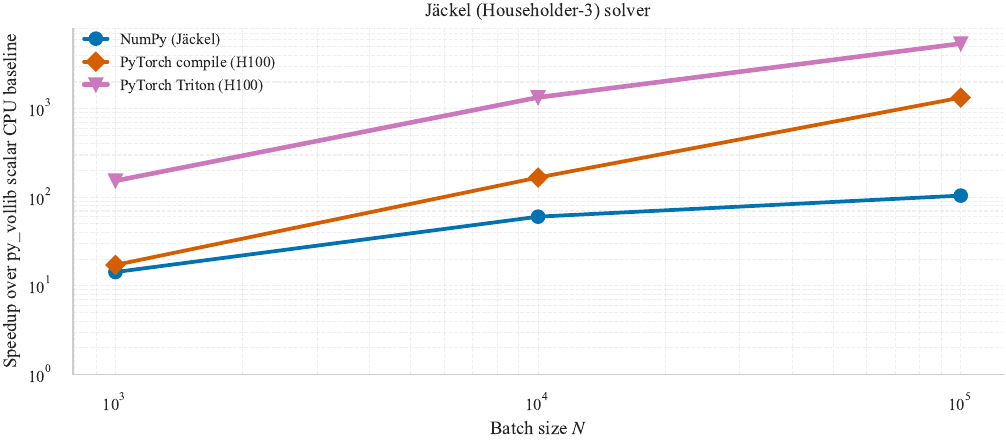}
\caption{J\"{a}ckel IV solver optimisation trajectory on $N=10^5$ options.
The CPU chain (NumPy $\to$ Numba) and GPU backends (\texttt{torch.compile},
JAX, Triton) are shown relative to the NumPy baseline of
\Cref{tab:jackel-bench}.}
\label{fig:iv-speedup-jackel}
\end{figure}

\paragraph{J\"{a}ckel solver, $N=10^5$ options.}
Table~\ref{tab:jackel-bench} summarises the optimisation trajectory of the
\fv\ J\"{a}ckel module on a fixed $N=10^5$ benchmark grid. CPU timings use a single thread
with NumPy / Numba; GPU timings use an NVIDIA H100 NVL with
CUDA-event-based measurement. ``Max rel.\ err.'' is measured against the
reference \texttt{py\_lets\_be\_rational} C implementation. The same trajectory
is plotted in \Cref{fig:iv-speedup-jackel}, which makes the per-stage gains
visually evident.

\begin{table}[!htb]
\centering\small
\setlength{\tabcolsep}{4pt}
\begin{tabular}{lrrl}
\toprule
\textbf{Stage / backend} & \textbf{Compute (ms)} & \textbf{Speedup} & \textbf{Max rel.\ err.} \\
\midrule
\multicolumn{4}{l}{\emph{CPU chain (NumPy $\to$ Numba)}} \\
NumPy baseline                & 106.5 & 1.0$\times$  & $3.2\!\times\!10^{-14}$ \\
+ fused vega                  &  88.9 & 1.2$\times$  & $3.2\!\times\!10^{-14}$ \\
+ Numba Householder           &  58.6 & 1.8$\times$  & $3.2\!\times\!10^{-14}$ \\
+ Numba boundary kernel       &  37.0 & 2.9$\times$  & $3.2\!\times\!10^{-14}$ \\
+ Hermite initial guess       &  15.5 & 6.9$\times$  & $1.7\!\times\!10^{-15}$ \\
+ Numba Hermite kernel        &  11.6 & 9.2$\times$  & $1.7\!\times\!10^{-15}$ \\
+ Numba pre/post-proc         &   8.5 & 12.5$\times$ & $2.2\!\times\!10^{-15}$ \\
\midrule
\multicolumn{4}{l}{\emph{GPU backends, H100 NVL}} \\
\texttt{torch.compile}        &  2.7  & 39$\times$   & $2.8\!\times\!10^{-15}$ \\
JAX \texttt{lax.fori\_loop}   &  2.4  & 44$\times$   & $4.9\!\times\!10^{-15}$ \\
Triton single-pass            &  0.056 & 1{,}900$\times$ & $9.3\!\times\!10^{-14}$ \\
\bottomrule
\end{tabular}
\vspace{0.5em}
\caption{\fv\ J\"{a}ckel IV solver, $N{=}10^5$ options. CPU timings: single-thread
NumPy/Numba. GPU timings: NVIDIA H100 NVL, CUDA-event compute time.}
\label{tab:jackel-bench}
\end{table}

\begin{figure}[t]
\centering
\includegraphics[width=0.95\linewidth]{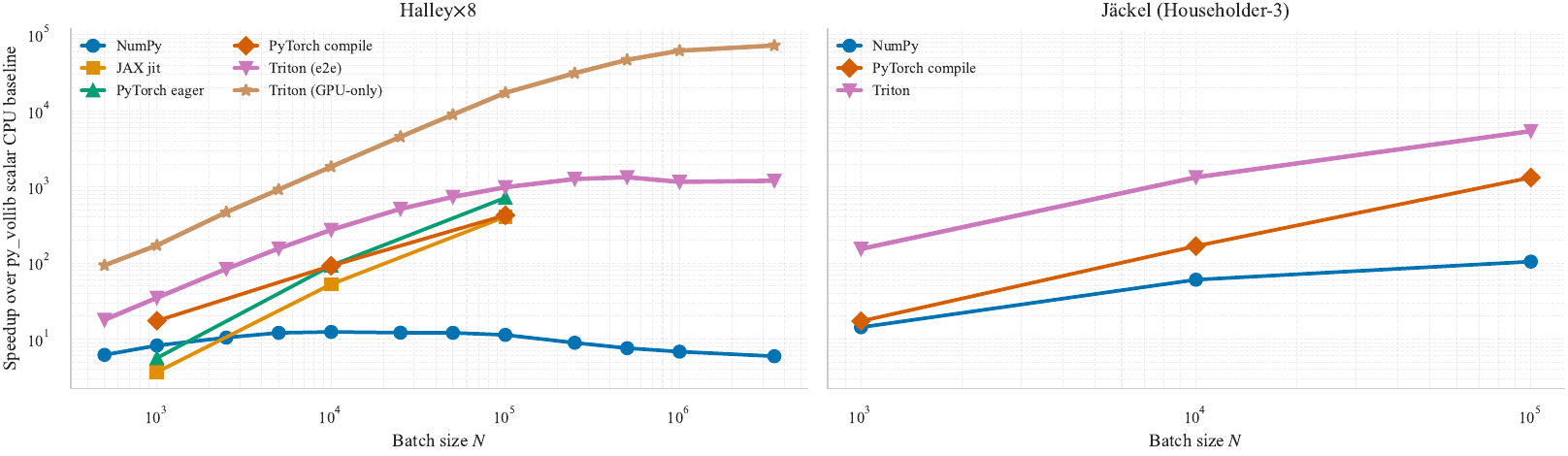}
\caption{End-to-end speedup of all \fv\ backends relative to the \pvv\ CPU
baseline of \Cref{tab:pvv-baseline}, including both Halley (\textbf{left}) and J\"{a}ckel (\textbf{right})
solver families.}
\label{fig:iv-speedup-all}
\end{figure}

The speedup column of \Cref{tab:jackel-bench} is computed against the CPU
NumPy baseline in the same table and is not a head-to-head comparison against
scalar \texttt{py\_lets\_be\_rational}. In our benchmark, all backends remain
within $\sim 10^{-13}$ relative error of that reference, i.e.\ at the
floating-point noise floor of \texttt{float64} IV inversion.
\Cref{fig:iv-speedup-all} situates the J\"{a}ckel stack alongside the Halley
backends against the legacy \pvv\ CPU reference.

\begin{figure}[t]
\centering
\includegraphics[width=0.85\linewidth]{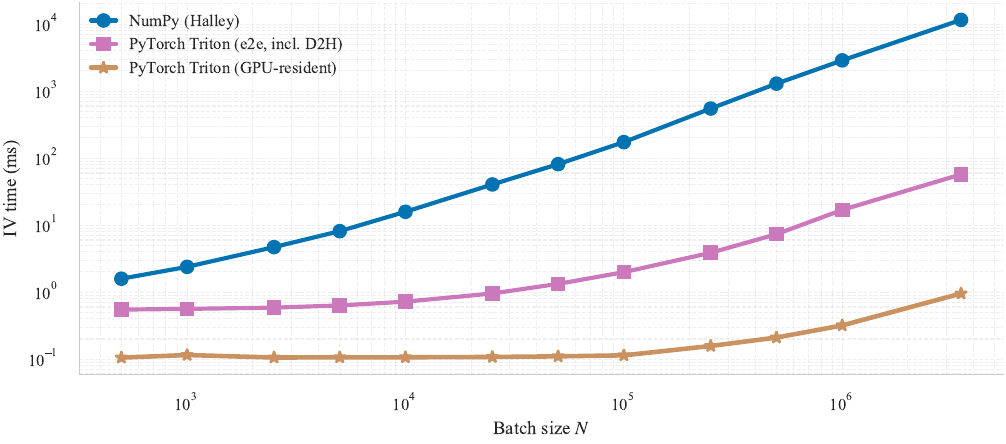}
\caption{IV throughput as a function of batch size $N$. CPU backends scale
near-linearly until cache effects appear; GPU backends amortise launch and
JIT cost above $N\!\sim\!10^4$.}
\label{fig:iv-scaling}
\end{figure}

\begin{figure}[t]
\centering
\includegraphics[width=0.95\linewidth]{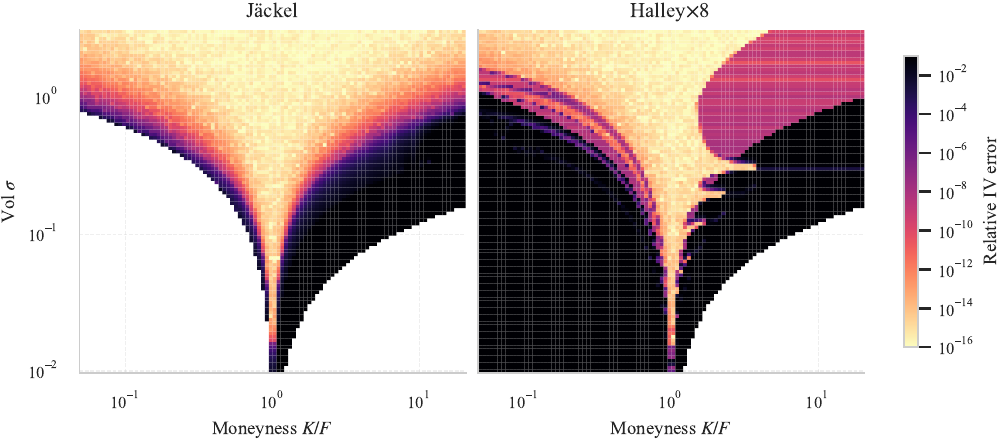}
\caption{Worst-case relative error of each \fv\ solver against the reference
\texttt{py\_lets\_be\_rational} C implementation, evaluated on a dense grid of
moneyness $\times$ total variance.}
\label{fig:solver-accuracy}
\end{figure}

\paragraph{Scaling with batch size and accuracy.}
\Cref{fig:iv-scaling} reports throughput as the number of options $N$ varies,
exposing the regimes in which CPU vectorisation dominates, where JIT compilers
amortise their warm-up cost, and where the Triton kernel saturates the GPU.
\Cref{fig:solver-accuracy} cross-checks the worst-case relative error of every
solver against the reference C implementation, confirming that the speedups
are not bought at the cost of precision: every backend stays within
single-ULP-class agreement on the bulk of the $(x,s)$ plane.

\section{Resources}\label{sec:resources}

\fv\ is open source under the MIT license. The canonical resources are:

\begin{itemize}\itemsep=1pt\parskip=0pt\topsep=2pt
\item \textbf{Source code (GitHub).}\\
      \url{https://github.com/raeidsaqur/fast-vollib}
\item \textbf{Stable releases (PyPI).}\\
      \url{https://pypi.org/project/fast-vollib/}
\item \textbf{Development snapshots (TestPyPI).}\\
      \url{https://test.pypi.org/project/fast-vollib/}
\item \textbf{Documentation.}\\
      \url{https://raeidsaqur.github.io/fast-vollib/}
\end{itemize}

Stable releases are tag-driven from \texttt{main}; \texttt{.devN} snapshots
are published from each commit on \texttt{main} to TestPyPI. Versioning is
VCS-derived via \texttt{hatch-vcs}.

\section{Limitations and Reproducibility}\label{sec:limits}

\fv\ targets European-style options under Black-76, Black-Scholes, and
Black-Scholes-Merton; American, exotic, and stochastic-volatility models are
out of scope and are better served by QuantLib~\citep{quantlib} or
model-specific libraries. The package requires Python~$\geq$~3.11; PyTorch and
JAX are optional and only needed for their respective backends. The Triton
J\"{a}ckel kernel requires a CUDA-capable PyTorch install. Quantitative
benchmark claims are intentionally left to the repository scripts and
documentation, where users can inspect the exact environment and re-measure on
their own hardware.

\paragraph{Conclusion.}
\fv\ packages batched Black-Scholes pricing, IV inversion, and Greeks behind a
small \pv-compatible Python API with NumPy, PyTorch, and JAX backends, plus an
experimental GPU-fused J\"{a}ckel solver. It targets ML/AI quant pipelines that
need batched, optionally differentiable, optionally GPU-accelerated IV
computation without breaking compatibility with the existing \pv\ ecosystem.
Contributions and bug reports are welcome via the GitHub repository.

\clearpage
{\small
\bibliographystyle{plainnat}
\bibliography{fv}
}

\end{document}